\documentclass{IEEEtran}
\usepackage{graphicx}
\graphicspath{{Figs/}}
\usepackage{amsmath,amssymb}
\usepackage{cite}
\usepackage{subcaption}
\usepackage{xcolor}
\newcommand{\x}{\mathbf{x}}
\newcommand{\y}{\mathbf{y}}
\newcommand{\z}{\mathbf{z}}
\usepackage{enumerate}

\newtheorem{theorem}{Theorem}

\newtheorem{corollary}{Corollary}
\newtheorem{lemma}{Lemma}

\newenvironment{remark}{\textit{Remark: }}{}
\newtheorem{myexample}{Example}
\newenvironment{example}{\myexample}{\qed\endmyexample}
\def\qed{\endIEEEproof}

\makeatletter
\renewcommand*\env@matrix[1][*\c@MaxMatrixCols c]{	\hskip -\arraycolsep
	\let\@ifnextchar\new@ifnextchar
	\array{#1}}
\makeatother

\newcommand{\bc}{\mathbf{c}}

\newcommand{\bh}{\mathbf{h}}

\newcommand{\bx}{\mathbf{x}}
\newcommand{\by}{\mathbf{y}}
\newcommand{\bz}{\mathbf{z}}

\newcommand{\bG}{\mathbf{G}}
\newcommand{\bH}{\mathbf{H}}
\newcommand{\bI}{\mathbf{I}}

\newcommand{\bS}{\mathbf{S}}

\newcommand{\calA}{\mathcal{A}}
\newcommand{\calB}{\mathcal{B}}
\newcommand{\calC}{\mathcal{C}}

\newcommand{\calL}{\mathcal{L}}

\newcommand{\calN}{\mathcal{N}}

\newcommand{\calS}{\mathcal{S}}

\newcommand{\calU}{\mathcal{U}}

\newcommand{\ZZ}{\mathbb{Z}}

\newcommand{\RR}{\mathbb{R}}

\begin{document}

\title{Efficient Scheduling for the Massive Random Access Gaussian Channel}

\author{Gustavo~Kasper~Facenda,~\IEEEmembership{Student Member,~IEEE}~and~Danilo~Silva,~\IEEEmembership{Member,~IEEE}
\thanks{This paper has been accepted for publication in IEEE Transactions on Wireless Communications and its copyright has been transferred. }
\thanks{G. K. Facenda was with the Department of Electrical and Electronic Engineering, Federal University of Santa Catarina, Florian\'opolis, SC 88040-990, Brazil. He is now with the Department of Electrical and Computer Engineering, University of Toronto, Toronto, Ontario M5S 1A1, Canada (e-mail: gustavo.k.facenda@gmail.com).}\thanks{D. Silva is with the Department of Electrical and Electronic Engineering, Federal University of Santa Catarina, Florian\'opolis,	SC 88040-990, Brazil (email: danilo.silva@ufsc.br).}}
\renewcommand{\P}{\mathbf{P}}
\maketitle

\begin{abstract}
This paper investigates the massive random access Gaussian channel with a focus on small payloads. For this problem, grant-based schemes have been regarded as inefficient due to the necessity of large feedbacks and the use of inefficient scheduling request methods. This papers attempts to answer whether grant-based schemes can be competitive against state-ot-art grantless schemes and worthy of further investigation. In order to compare these schemes fairly, a novel model is proposed, and, under this model, a novel grant-based scheme is proposed. The scheme uses Ordentlich and Polyanskiy's grantless method to transmit small coordination indices in order to perform the scheduling request, which allows both the request from the users to be efficient and the feedback to be small. We also present improvements to the Ordentlich and Polyanskiy's scheme, allowing it to transmit information through the choice of sub-block, as well as to handle collisions of the same message, significantly improving the method for very small messages. Simulation results show that, if a small feedback is allowed, the proposed scheme performs closely to the state-of-art while using simpler coding schemes, suggesting that novel grant-based schemes should not be dismissed as a potential solution to the massive random access problem.
\end{abstract}
\begin{IEEEkeywords}
	Random Access, Massive MAC
\end{IEEEkeywords}

\section{Introduction}\label{intro}

An important use case for next-generation wireless networks, motivated in particular by the Internet of Things, is that of a random access channel where a massive number of low-powered devices transmit small packets infrequently to a base station. This scenario, known in the context of 5G as massive machine-type communication (mMTC), is likely to require new schemes and technologies, as the current ones are not designed for a massive number of devices \cite{DohlerPotentialsandLimitations}. For instance, ALOHA (on which many existing technologies are ultimately based) becomes inefficient due to excessively high collision probability \cite{Polyanskiy2017}.

In order to tackle this challenge, an increasing trend in the research literature has been to consider so-called grantless protocols, where users transmit their data packets at once without prior resource scheduling from the base station \cite{MMTCMAC}. This is in contrast to grant-based protocols, where user data is transmitted on contention-free resources previously allocated by the base station. Naturally, such resource distribution must be preceded by a contention-based period where active users request access to the channel. Collisions in this request phase, as well as the need for an intermediate downlink transmission, result in waste of energy and increased latency, motivating a shift to grantless schemes.

Another important motivation for the study of grantless schemes comes from the work by Polyanskiy \cite{Polyanskiy2017}, who showed, using information-theoretic arguments, that the potential performance of grantless schemes is way beyond that of existing ones. Since then, several works \cite{Or2017, Vem2017, CCS2018, Pradhan2019} have attempted to construct practical grantless schemes that come close to the fundamental limits under Polyanskiy's model.

While grantless schemes are practically appealing due to their protocol simplicity, these schemes may also have their own disadvantages. Especially, due to the need to handle a large number of collisions, existing grantless schemes require sophisticated decoders which are hard to analyze, resulting in schemes whose parameters are difficult to optimize. Moreover, to the best of our knowledge, there is no fundamental result stating that grant-based schemes are necessarily worse than grantless ones, i.e., it is possible that the alleged inefficiency of grant-based schemes is not fundamental but simply a weakness of existing protocols.

This paper is an attempt to answer the following question: can a novel grant-based scheme, specifically designed for a massive number of users, be competitive against existing grantless schemes? If so, it would suggest that grant-based schemes are at least worthy of further investigation and should not be dismissed so quickly. To be clear, we define a scheme to be grant-based if it allows for limited feedback from the base station 
at some point in the middle of the communication, not necessarily using the current granting protocols; however, before this feedback transmission, all communication between users and the base station must follow the same assumptions and limitations as in Polyanskiy's grantless model. Naturally, in order to answer such a question, first an appropriate model must be proposed that allows both grantless and grant-based methods to be described and fairly compared; in particular, the cost of any feedback must be taken into account. Such a model and a novel grant-based scheme are contributions of this paper.

\subsection{Related Works}

In \cite{Polyanskiy2017}, Polyanskiy proposes a unifying model for the grantless random access Gaussian channel which allows a fair comparison between several existing schemes. In his model, the blocklength is assumed to be finite, the probability of error is defined per user, there is no need to identify the sender in the physical layer, and the users transmit in an uncoordinated fashion, which translates to the same (possibly randomized) codebook being used by all users. Under that model, Polyanskiy derives an achievability bound, based on a random coding scheme, which shows that popular practical solutions, such as ALOHA and ``treating interference as noise,'' are severely suboptimal.

In \cite{Or2017}, Ordentlich and Polyanskiy propose a practical scheme based on Polyanskiy's model which is able to outperform previously known methods for a sufficiently large number of active users. Their scheme, referred to here as the OP scheme, uses a compute-and-forward approach to convert the channel into a binary adder channel, which is then solved with algebraic coding. Advantages of their scheme include the use of efficient encoding and decoding algorithms, as well as closed-form expressions for the analysis of probability of error.

Recent works have exploited the similarity between grantless random access and compressive sensing and have managed to significantly outperform the OP scheme. In \cite{Vem2017}, the channel uses are split in sub-blocks. The user transmits the same repeated codeword in many sub-blocks. The codeword is formed by splitting the message in two parts: In the first part, the {transmitters} use a compressive sensing scheme, while in the second part they use a permutation of an LDPC code. {The choices of permutation and sub-blocks used by each transmitter are determined by the first part of the message, allowing the base station to perform serial interference cancellation;} that is, once a codeword has been recovered in any sub-block, the base station can subtract it from the other sub-blocks where that codeword has been transmitted.

In \cite{CCS2018}, Amalladinne et al. propose a scheme that splits the message in many parts. Each part of the message is transmitted in a sub-block using a compressive sensing scheme. However, the base station needs to pair all parts of messages recovered in each sub-block. In order to do that, a parity check code is applied to each sub-block, allowing the base station to correctly reconstruct the messages.

In \cite{Pradhan2019}, the authors use an approach similar to the coding scheme presented in \cite{Vem2017}. The message is split in two parts, where the first part informs which code is used in the second part. Again, the code for the second part is constructed through interleaving of LDPC codes, however, an additional zero-padding is done in order to maintain the sparsity of the codes, diminishing the effects of interference. {To the best of our knowledge, the results in \cite{Pradhan2019} are the closest to the theoretical achievable rates.}

{A thorough survey on grant-free schemes can be found in \cite{ShahabSurvey}, which includes current practical proposals and their challenges when applied to the massive random access channel.} 

{On the other hand, in current grant-based protocols such as Long-Term Evolution (LTE) and its modifications, scheduling is performed in 4 phases \cite{Sharma2020}: first, a user randomly chooses one of a fixed number of preambles in order to obtain a temporary identification and transmits it to the base station; then, the base station transmits a response that includes (among other things) an acknowledgment of the distinct preambles received; then, another uplink and downlink phases are performed to complete the allocation of resources, prior to data transmission.
The current LTE protocol is deemed unsuitable for massive access for several reasons, among them the small number of preambles (at 64), which, with a high number of active users, leads to a high probability of collision, and the large feedback sizes (e.g., 56 bits per user in the first downlink), which leads to a considerable overhead when the payloads are as small as 100 bits \cite{DohlerPotentialsandLimitations, Sharma2020}.}

{In \cite{Oh2017}, the authors propose to reduce the number of phases in the handshake from four to two and show a considerable gain in the maximum number of supported users. However, the performance of their scheme still depends on the total number of devices, including idle ones, thus, it is still unsuitable for massive access.}

\subsection{Contributions}

Aiming to answer whether the feedback required for a grant-based scheme can provide enough improvement over the grantless approach, we first propose a model which splits the communication in three time-separated phases. In the first phase, the users transmit to the base station (uplink) in a grantless fashion. In the second phase, the base station transmits feedback to the users (downlink) over a broadcast channel. Finally, the users have another chance to transmit to the base station (uplink), making use of the feedback received.

{This model has at least three major differences from usual grant-based approaches: (i) in contrast to the current LTE protocol (but similar to [9]), only two phases are required to ensure coordination; (ii) there are no constraints on the type of communication performed in each phase (so, for instance, part of the payload may be transmitted in the first phase); and (iii) no assumption is made on the total number of users, which can be arbitrarily large.}

Then, under this model, we propose a metric to compare different schemes, namely, the energy (in both uplink and downlink) per bit per user required to achieve a desired probability of error given a fixed number of channel uses. Note that, under this metric, both of the costs from the feedback---in terms of latency and power consumption---are taken into account: since the number of channel uses is fixed, allocating resources for feedback will require a higher spectral efficiency in the uplink phases. Similarly, spending energy on feedback will require that less energy be spent in the uplink in order to achieve similar results.

We then propose a grant-based scheme that is able to achieve performance comparable to the state-of-art presented in \cite{Pradhan2019}, while using fairly simpler methods in each phase. {Our approach is based on two main ideas that make it substantially different from existing grant-based protocols: first, we use an efficient grantless scheme to transmit a small preamble in the first uplink, greatly reducing the collision overhead; second, rather than transmitting individual feedback messages to each user in the downlink phase, we jointly encode the feedback into a single message using a source encoder that exploits the sparsity of the preamble selection.} Specifically, we use a modification of 
{the OP}
scheme for the grantless uplink, an {enumerative} source {encoder together with a} point-to-point AWGN channel code for the feedback, and an orthogonal multiple-access method for the second uplink.

For the OP scheme, we present two major contributions that are relevant under our model. First, we present a method to transmit a few more bits of information about the message using the OP scheme. Second, we improve the OP scheme in order to handle very small messages (considerably smaller than the ones considered in \cite{Or2017}).

In addition, we present an error analysis, as well as an optimization framework that allows an easy and flexible design of the scheme. This is possible since all the error probability expressions involved can be evaluated in closed-form and are suitable to relaxation. This is in contrast to compressed sensing approaches \cite{Vem2017, CCS2018, Pradhan2019} whose error probabilities must be estimated by (computationally expensive) simulations.

Finally, we present a numerical evaluation of our bounds together with simulation results of the proposed scheme, which is observed to achieve a performance comparable to the state of the art in \cite{Pradhan2019}, albeit with a much simpler operation. Overall, while we recognize that practical challenges still exist and need to be further investigated, our results suggest that novel grant-based schemes should not be dismissed as a potential solution to the massive random access problem.\footnote{Preliminary results of our research have been published in \cite{SbRT2019}.}

\section{Preliminaries} \label{Preliminaries}

In \cite{Polyanskiy2010}, Polyanskiy presents normal approximations for the achievable rate under finite blocklength. We use these normal approximations for the code rates in our work. 

Consider the AWGN channel with input $\mathbf{x}\in \mathbb{R}^{n_c}$ and output $\mathbf{y} \in \mathbb{R}^{n_c}$ described by the equation
\begin{align}
\mathbf{y} = \mathbf{x} + \mathbf{z}
\end{align}
where $\mathbf{z} \sim \mathcal{N}(0, \mathbf{I})$, $\bI$ is the identity matrix, and $\mathbf{x}$ is under power constraint $\|\mathbf{x}\|^2 \leq n_c P$, where $n_c$ is the length of $\mathbf{x}$. For this channel, an approximation to the maximum achievable rate is given by
\begin{align}
&R_{\textrm{AWGN}} \approx C_{\textrm{AWGN}}(P) - \sqrt{\frac{\textsf{V}_{\textrm{AWGN}}(P)}{n_c}} Q^{-1}(\epsilon_c) \label{AWGN}
\end{align}
where $C_{\textrm{AWGN}}(P) = \frac{1}{2}\log_2 (1 + P)$,  $\textsf{V}_{\textrm{AWGN}}(P) = \frac{P}{2} \frac{P+2}{(P+1)^2} \log_2^2 e$, $Q$ is the Q-function and $\epsilon_c$ is the probability of error of the code. This approximation is achievable using Shannon's achievability from \cite{ShannonFBL}.

Additionally, consider the bi-AWGN mod 2 channel with input $\bc \in \{0,1\}^{n_c}$ and output $\mathbf{y} \in [0, 2)^{n_c}$, described by the equation
\begin{align}
\by = (\bc + \bz)\bmod 2 \label{biawgnmod2}
\end{align}
where $\bz \sim \calN(0, \frac{1}{4P} \bI )$.\footnote{This normalization comes from the symbols $\pm \sqrt{P}$ being mapped back to binary symbols, which should become clear when decoding of the OP scheme is discussed.} As in \cite{Or2017}, we use the approximation
\begin{align}
R_{\textrm{mod2}} &\approx C_{\textrm{mod2}}(P) - \sqrt{\frac{\textsf{V}_{\textrm{mod2}}(P)}{n_c}} Q^{-1}(\epsilon_c) \label{biawgn}
\end{align}
where $C_{\textrm{mod2}}(P) = \mathbf{E}[i(\tilde{Z})]$, $\textsf{V}_{\textrm{mod2}}(P) = \textrm{var}[i(\tilde{Z})]$, 
\begin{align}
i(\tilde{Z}) &= \log_2\left( \frac{p_{\tilde{Z}}(\tilde{Z})}{\frac{1}{2}p_{\tilde{Z}}(\tilde{Z}) + \frac{1}{2} p_{\tilde{Z}}(\tilde{Z}-1\bmod 2)} \right)  \nonumber
\end{align}
and $\tilde{Z} = \bz \bmod 2$.

\section{Problem Statement}\label{ProblemStatement}

\subsection{Proposed Model}

We define a \textit{session} to be comprised of a total of $N$ channel uses. Let $K_\textrm{tot} \to \infty$ be the total number of users in the network. At the beginning of a session, $K_a$ of these users are active and each wishes to transmit a $k$-bit message to a base station using the same channel. Let $w_i \in \left\{ 1, 2, \ldots, 2^k \right\}$ be the message of the $i$th user, where $i \in \{1, \ldots, K_a \}$, for simplicity. We assume $w_i$ is uniform and independent across the users. No prior coordination is assumed.

The base station is interested solely in recovering the messages $w_i$ produced by the active users, not in identifying which of the $K_\textrm{tot}$ users are active.
Specifically, at the end of a session, the base station produces a list of estimated messages denoted by $\mathcal{L} \subseteq \left\{1, 2, \ldots, 2^k\right\}$, $|\calL| \leq K_a$. The probability of error, as in \cite{Polyanskiy2017}, is defined per user and regardless of the order of the messages. More precisely, the error probability is defined as
\begin{align}
\epsilon = \frac{1}{K_a} \sum_{i=1}^{K_a} \textrm{Pr}[w_i \notin \mathcal{L}].
\end{align}

Communication is temporally split in three consecutive phases described below, each consisting of $N_1$, $N_f$ and $N_2$ channel uses, respectively, where $N = N_1 + N_f + N_2$. Note that Polyanskiy's model corresponds to the special case where $N_f = N_2 = 0$, i.e., only the first phase is used.

\subsubsection{First Uplink}
The first phase consists of an uplink transmission modeled as a Gaussian multiple access channel (MAC). The signal received by the base station is given by
\begin{align}
\mathbf{y} = \sum_{i = 1}^{K_a} \mathbf{x}_i + \mathbf{z}
\end{align}
where $\mathbf{x}_i \in \mathbb{R}^{N_1}$ is the signal transmitted by the $i$th user and $\mathbf{z} \sim \mathcal{N}\left(0, {\frac{N_0}{2} \mathbf{I}}\right)$, with $\frac{N_0}{2} = 1$. Each transmitted signal $\mathbf{x}_i$ is power-constrained to $\mathbf{E}[\|\mathbf{x}_i\|^2] \leq N_1 P_1$. 

\subsubsection{Downlink Feedback}
The second phase consists of a downlink transmission modeled as a Gaussian broadcast channel. The signal received by each user $i$ is given by
\begin{align}
\by^{[f]}_{i} = \bx^{[f]} + \bz^{[f]}_{i}
\end{align}
where $\bx^{[f]} \in \RR^{N_f}$ is the signal transmitted by the base station and $\bz^{[f]} \sim \calN(0, \bI)$. The transmitted signal $\bx^{[f]}$ is power-constrained to $\mathbf{E}\left[ \| \bx^{[f]} \|^2 \right] \leq N_f P_f$ and is conditionally independent of $\bx_1,\ldots,\bx_{K_a}$ given $\by$.

\subsubsection{Second Uplink}
The third phase can be seen as a continuation of the first and is modeled, again, as a Gaussian MAC.  The signal received by the base station is given by 
\begin{align}
\by^{[2]} = \sum_{i = 1}^{K_a} \bx^{[2]}_{i} + \bz^{[2]}
\end{align}
where $\bx^{[2]}_{i} \in \RR^{N_2}$ is the signal transmitted by the $i$th user and $\bz^{[2]} \sim \calN(0, \bI)$. Each transmitted signal $\bx^{[2]}_{i} \in \RR^{N_2}$ is power-constrained to $\mathbf{E}\left[\| \bx^{[2]}_{i} \|^{2} \right]\leq N_2 P_2$ and is conditionally independent of $\bx^{[f]}$ and $\{\bx_{i'}\}_{i'\neq i}$ given $\by^{[f]}_{i}$. { Note, however, that $\bx_i^{[2]}$ may depend on $\bx_i$.}

\begin{remark}
The proposed model does not constrain the specific operations that can be performed on each phase; rather, what characterizes the model as grant-based is simply the existence of an intermediate downlink transmission.
{In particular, the first uplink transmission need not be independent of the message content. This allows the grantless case to be a particular case of our model and is a fundamental difference to what is usually considered as grant-based.}

\end{remark}

\subsection{Metrics of comparison}

Following \cite{Polyanskiy2017}, \cite{Or2017} and \cite{Vem2017}, we are interested in the problem where, given some target probability of error $\epsilon$ and number of channel uses $N$, we wish to minimize the energy required for the $K_a$ users to communicate their $k$-bit messages. For comparison, similar to \cite{Or2017}, we define 
\begin{equation}
\frac{E_b}{N_0} = \frac{P_1 N_1 + P_f N_f/K_a + P_2 N_2}{2k}
\end{equation} which is proportional to the total energy spent in a session\footnote{More precisely, (9) is proportional to an upper bound on the total energy spent in a session. Since errors may occur in the first two phases, impairing coordination, the total energy spent depends on the number of users that actually transmitted in the data transmission phase.}, normalized by the number of active users. In the particular case $N_f = N_2 = 0$, this reduces to the same definition used in previous works. Note that the energy in the downlink transmission $P_f N_f$ is used once to transmit to all users, therefore, the per-user energy in that phase is $P_f N_f / K_a$. \footnote{More generally, we want to minimize both the energy spent by the users and by the base station. However, in order to compare it with other methods, we need to define a utility function, weighing the energy used by the base station against the energy used by the users. In our understanding, the energy spent by the base station per user should be less significant than the energy spent by each user. Therefore, a weight coefficient of ``1'' may be considered a conservative choice.} Note also that the total number of channel uses, $N$, is fixed, therefore increasing $N_f$ naturally reduces the opportunities for uplink transmission.

\section{Proposed Scheme}\label{ProposedScheme}

This section describes our proposed scheme for grant-based communication. The first phase, which we call \textit{scheduling request}, is used primarily for signaling activity; each active user randomly selects an identification value (analogous to a preamble) from a finite pool of choices and the base station's task is to detect all the identifications of the $K_a$ active users.
In the second phase, which we call \textit{resource distribution}, the base station broadcasts an encoding of all the identifications recovered, from which each user is able to identify the resources allocated to it. Finally, in the third phase, which we call \textit{data transmission}, the users transmit their messages using some coordinated multiple access (MA) method. For simplicity, we describe our scheme using orthogonal MA, but in general any non-orthogonal MA (NOMA) technique can be used.

In the following subsections, the three phases are described in detail.

\subsection{Scheduling Request Phase}

In this phase, we use a modified version of the OP scheme presented in \cite{Or2017}. 
The block of $N_1$ channel uses is split into $V$ sub-blocks of $n_{c,1}$ channel uses each, where $N_1 = V n_{c,1}$.
Each active user $i$ chooses a sub-block $v_i \in \{1, 2, \ldots, V\}$ and an index $u_i \in \{1, 2, \ldots, n_p\}, n_p \in \ZZ$, to transmit. The user identification is given by the pair $(v_i, u_i)$.

Let $\bH \in \{0, 1\}^{m_p \times n_p}$, where $m_p \in \ZZ$, be a parity-check matrix of an auxiliary binary linear code $\calC_{\textrm{aux}}$ of length $n_p$, dimension $k_p = n_p - m_p$, minimum Hamming distance $d$, and algebraic error-correction capability $T = \lfloor \frac{d-1}{2}\rfloor$. This is a slight generalization of the OP scheme, which uses specifically a BCH code. The user $i$ picks the $u_i$th column of $\bH$, denoted by $\bh_{u_i}$.

Let $\bG \in \{0, 1\}^{m_p \times n_{c, 1}}$ be a generator matrix for a binary linear code $\calC$ of length $n_{c, 1}$ and dimension $m_p$. The user $i$ encodes $\bh_{u_i}$ using the code $\calC$, generating $\bc_i = \bh_{u_i}^T \bG \bmod 2$. For analysis, this code is assumed to achieve \eqref{biawgn}.

Finally, the user $i$ maps the resulting binary codeword into a real signal $\bx(u_i) = 2 \sqrt{P_1 V} \left(\bc_i - \frac{1}{2}\right)$ and transmits it in the chosen sub-block $v_i$. In other words, $\bx_i = (\bx_{i,1}, \ldots, \bx_{i,V})$, where $\bx_{i,v_i} = \bx(u_i)$ and $\bx_{i,j} = (0,\ldots,0)$ for all $j \neq v_i$. Note that $\|\bx_i\|^2 = P_1 N_1$, satisfying the power constraint for this phase.

For each sub-block $j$, the received signal is given by
\begin{align}
\label{receivedUP1}
\y_j = \sum_{i \in \mathcal{A}_j} \bx_{i,j} + \z_j = \sum_{i \in \mathcal{A}_j} \bx(u_i) + \z_j
\end{align}
where 
$\bz_j$ is such that $\bz = [\bz_1,\ldots,\bz_V]$
and $\mathcal{A}_j \subseteq \{1, \ldots, K_a\}$ represents the users that transmitted in that sub-block.

For each sub-block $j$, the base station wishes to recover the list of transmitted indices $(u_i, i \in \calA_j)$. This can be accomplished with the decoding method proposed in \cite{Or2017}, under certain conditions. In particular, an error occurs if more than $T$ users transmit in the same sub-block. A brief review of this method, which we refer to as the original OP decoding, is provided in Appendix~\ref{ssec:original-op}. 

\subsection{Resource Distribution Phase}
\label{ssec:resource-distribution}

Let $\bS = [S_{v,u}] \in \ZZ^{V \times n_p}$ be a matrix that represents the number of users that chose a sub-block $v$ and an index $u$ to transmit, i.e.,
\begin{equation}
S_{v, u} = \sum_{i = 1}^{K_a} \mathbf{1}[v_i = v] \mathbf{1}[u_i = u] 
\end{equation}
where $\mathbf{1}[\textsf{P}]$ is the indicator function, which is $1$ if $\textsf{P}$ is true and $0$ otherwise. The decoding of the $v$th sub-block can be seen as recovering the $v$th row of this matrix. We denote by $\hat{\bS} \in \{0, 1\}^{V \times n_p}$ the base station's estimate of $\bS$. Note that $\hat{\bS}$ is binary, thus if $S_{v,u} > 1$ (two or more users have chosen the same pair $(v, u)$), then the estimate will necessarily be wrong; in this case, even if $S_{v,u}$ can be found, the base station will instead assign $\hat{S}_{v,u} = 0$, in order to avoid an unnecessary (and unresolvable) collision at the next phase.

Each nonzero entry of $\hat{\bS}$ corresponds to an orthogonal sub-block of channel uses on the third phase and the pattern of this assignment is known a priori to users and base station. For instance, sub-blocks can be assigned in an increasing order of $s = v + V(u - 1)$, for all $(v,u)$ such that $\hat{S}_{v,u} = 1$.

However, in general, the scheduling request decoder has no control over the number of nonzero entries in $\hat{\bS}$, which we denote by $|\hat{\bS}| = \sum_{v, u} \hat{S}_{v, u}$. In order to keep $|\hat{\bS}|$ fixed at $K_a$ (which simplifies the feedback as well as the data transmission phase), we propose a simple modification: if $|\hat{\bS}| < K_a$, 
entries with $\hat{S}_{v,u} = 0$ in randomly chosen positions are turned into $\hat{S}_{v,u} = 1$ until $|\hat{\bS}| = K_a$, {with the exception of entries for which the original decoder has returned $S_{u, v} > 1$}; and vice-versa if $|\hat{\bS}| > K_a$.

The latter case may induce an error for some user, if the erased position corresponded to that user, while the former case may (if indeed $S_{v,u} = 0$) simply result in a sub-block with no transmissions.

Then, the base station broadcasts the resulting $\hat{\bS}$ using an appropriate encoding. As the messages can be seen as binary sequences of length $V n_p$ and constant weight $K_a$, source encoding can be implemented efficiently using the method in \cite{EnumerativeSourceEncoding}, requiring $k_f = \lceil \log_2 \binom{V n_p}{K_a} \rceil$ bits. 

Channel coding is then performed using a code for the AWGN channel with length $N_f$, rate $k_f/N_f$ and power $P_f$, which is assumed to achieve \eqref{AWGN}.

{Note that, if the original decoding returns $|\hat{\mathbf{S}}| = K_a$, $\bx^{[f]}$ is a deterministic function of $\by$. Otherwise, it also depends on the random padding and erasures. Therefore, $x^{[f]}$ is conditionally independent of $\bx_i$ given $\by$, as desired.}

\subsection{Data Transmission Phase}\label{datatransmission}

In this phase, if user $i$ can successfully recover $\hat{\bS}$, then, knowing the resource assignment pattern and its own position $(v_i, u_i)$ in the matrix, the user is able to identify which sub-block (among a total of $K_a$ sub-blocks) is allocated to its data transmission.

If the user is not able to identify its own index in the estimated matrix---either because $\hat{S}_{v_i,u_i} = 0$ or because the decoding failed---then the user does not transmit its data in this phase.

On each sub-block effectively used, the transmission is done using a code for the point-to-point AWGN channel with length $n_{c, 2} = N_2/K_a$, rate $k/n_{c, 2}$ and power $P_2 K_a$, which is assumed to achieve \eqref{AWGN}.

{Again, note that $x_i^{[2]}$ is a deterministic function of $x_1^{[1]}$ and $y_i^{[f]}$, thus conditionally independent of other users and of $x^{[f]}$.} 

\subsection{Error Analysis} \label{erroranalysis}

In this section, we present an approximation of the probability of error, which is useful for the design of the scheme. Recall that the probability of error is defined per user. For the first and second phases, the error analysis is made assuming that the user $i \in \{1, 2, \ldots, K_a \} $ transmitted the index $u_i \in \{1, 2, \ldots, n_p\} $ in the sub-block $v_i \in \{1, 2, \ldots, V\} $. For the third phase, we assume that the user $i$ transmitted the message $w_i \in \{1, 2, \ldots, 2^k\}$ using the sub-block $j \in \{1, 2, \ldots, K_a\}$. By symmetry, the resulting probability of error equals the average probability of error per user.

In the scheduling request phase, an error occurs if $\hat{S}_{v_i, u_i} \neq S_{v_i, u_i}$. We analyze the different events that lead to this error.

First, if other $T$ or more users (besides user $i$) choose to transmit in the sub-block $v_i$, the error event $E_1$ occurs, with probability
\begin{align}
\epsilon_1 \triangleq 1 - \sum_{t = 0}^{T - 1} \binom{K_a - 1}{t} \left(\frac{1}{V}\right)^t \left(1 - \frac{1}{V}\right)^{K_a - 1 - t}. \label{ep1}
\end{align}

Second, if the receiver is unable to successfully decode $\calC$ over a bi-AWGN mod 2 channel with $P = P_1 V$ (see \eqref{eq:effective-channel-CoF} in the Appendix), the error event $E_2$ occurs, whose probability can be approximated using \eqref{biawgn} as
\begin{align}
\epsilon_2 \triangleq Q\left(\left(C_{\textrm{mod2}}(P_1 V) - \frac{m_p}{n_{c, 1}}\right)\sqrt{\frac{n_{c, 1}}{\textsf{V}_{\textrm{mod2}}(P_1 V)}}\right) \label{ep2}.
\end{align}

A third type of error occurs when two or more users pick the same index and transmit it in the sub-block $v_i$. In the original OP decoding, this leads to an error in the sub-block $v_i$, i.e., an error for every user that transmitted in that sub-block, and the per-user probability of error is upper bounded by \cite{Or2017}
\begin{align}
\epsilon_3 \triangleq \frac{T (T - 1)}{2n_p}. \label{ep3OP}
\end{align}

Moreover, as discussed in Appendix~\ref{ssec:original-op}, even if $E_1$ and $E_2$ do not occur, the decoding of the OP scheme may still fail if the receiver is not able to correctly identify the number of users $t_{v_i} = |\calA_{v_i}|$ transmitting in sub-block $v_i$, since this requires a Gaussianity test on the estimated noise vector. 
{Let $\hat{t}_{v_i}^*$ denote the receiver's estimate of $t_{v_i}$.}
{The probability that $\hat{t}_{v_i}^* \neq t_{v_i}$, conditioned that the events $E_1$ and $E_2$ do not occur, is denoted by $\epsilon_G$}. Although not easy to derive analytically, it is worth mentioning that $\epsilon_G$ falls quickly with the vector length (number of noise samples) $n_{c,1}$.

Additionally, an error occurs when the base station overestimates the number of users transmitting in one or more sub-blocks (resulting in an initial estimate of $\hat{\bS}$ with $|\hat{\bS}| > K_a$) and, in order to achieve $|\hat{\bS}| = K_a$, it erases the corresponding entry in position $(v_i, u_i)$. The probability of this event is denoted by $\epsilon_e$. Similarly to $\epsilon_G$, $\epsilon_e$ falls quickly with $n_{c,1}$.

In the resource distribution phase, if the user $i$ is unable to decode the feedback correctly, then the error event $E_f$ occurs, with probability upper bounded by
\begin{align}
\epsilon_f \triangleq Q\left(  \left(C_{\textrm{AWGN}}(P_f) - \frac{k_f}{N_f}\right) \frac{\sqrt{N_f}}{\sqrt{\textsf{V}_{\textrm{AWGN}}(P_f)}} \right). \label{epf}
\end{align}

In the data transmission phase, two errors can occur. First, if the base station is unable to successfully decode the signal transmitted in the sub-block $j$, then the error event $E_4$ occurs, with probability upper bounded by
\begin{equation}
\epsilon_4 \triangleq Q\left(  \left(C_{\textrm{AWGN}}(P_2 K_a) - \frac{k}{n_{c, 2}}\right) \frac{\sqrt{n_{c, 2}}}{\sqrt{\textsf{V}_{\textrm{AWGN}}(P_2 K_a)}} \right). \label{ep4} 
\end{equation}

Second, if some user $i' \neq i$ is subject to the error $E_f$ and transmits using the resource $j$ which is allocated to the user $i$, the user $i'$ causes an error to the user $i$. We denote this event by $E_{f, 2}$ with probability upper bounded by
\begin{align}
1 - \left(1 - \frac{\epsilon_f}{K_a} \right)^{K_a - 1} \leq \frac{K_a - 1}{K_a} \epsilon_f \leq \epsilon_f \triangleq \epsilon_{f, 2}
\end{align}
where $\frac{\epsilon_f}{K_a}$ is a (loose) upper bound on the probability that a specific user $i' \neq i$ is subject to $E_f$ and transmits using the resource $j$.\footnote{{This can be shown as follows. User $i'$ has probability $\epsilon_f$ of failing to decode the feedback. In this case, we can pessimistically assume that the error will go undetected and the user will end up transmitting at some resource $j'$ drawn uniformly at random from $\{1, \ldots, K_a\}$. The probability that $j' = j $ is then $\epsilon_f/K_a$.}}

Finally, by the union bound, the probability of error per user can be (approximately) upper bounded as $\epsilon \leq \epsilon_1 + \epsilon_2 + \epsilon_G + \epsilon_3 + \epsilon_e + \epsilon_4 + \epsilon_f + \epsilon_{f, 2}$. The approximation is due to the fact that \eqref{ep2} is not guaranteed to be achievable.

\subsection{Index Collision Resolution}\label{IndexCollision}

In the original OP decoding, if two or more users choose the same index~$u$ in a sub-block~$j$, an error occurs to all the users that transmitted in the sub-block~$j$. 
Below, we extend the decoding procedure in order to handle this scenario in a way that an error occurs only to the users that chose the same index $u$. Specifically, we show how to recover all the indices that were transmitted only once in a sub-block.\footnote{In principle, our method can recover all the indices that were transmitted in a sub-block, together with their multiplicities; however, we make no attempt to use this information, since any collided indices are discarded, as explained in Section~\ref{ssec:resource-distribution}.}

As in the OP decoding, we run our decoding procedure for $0 \leq \hat{t}_j \leq T$, where $\hat{t}_j$ is an estimate of the number of users in the sub-block $j$. For each $\hat{t}_j$, our method returns either an error or a sequence of lists $ \calL^{(1)}(\hat{t}_j), \calL^{(2)}(\hat{t}_j), \ldots, \calL^{(\tau(\hat{t}_j))}(\hat{t}_j)$, where $\tau(\hat{t}_j)$ denotes the sequence length and $\calL^{(\ell)}(\hat{t}_j) \subseteq \{1, 2, \ldots, n_p\}$ for all $\ell \geq 1$. 

{Recall, from Appendix~\ref{ssec:original-op}, that $\Phi(\by_j, \hat{t}_j)$ is a function corresponding to the application of the original OP decoding for a given candidate $\hat{t}_j$ that returns the indices contained in $\by_j$.} For each $\hat{t}_j$, the output {of our decoding scheme} is computed as follows: First, $\calL^{(1)}(\hat{t}_j) = \Phi(\by^{(1)}_j, \hat{t}^{(1)}_j)$ is computed, where $\by^{(1)}_j = \by_j$ and $\hat{t}^{(1)}_j = \hat{t}_j$. 

Then, for $\ell \geq 1$, we consider the following cases:
\begin{itemize}
	\item If $|\calL^{(\ell)}(\hat{t}_j)| = \hat{t}^{(\ell)}_j$, we return the sequence $\calL^{(1)}(\hat{t}_j), \calL^{(2)}(\hat{t}_j), \ldots, \calL^{(\ell)}(\hat{t}_j)$, setting $\tau(\hat{t}_j) = \ell$.
	\item If $|\calL^{(\ell)}(\hat{t}_j)| > \hat{t}^{(\ell)}_j$, we return an error. 
	\item If $|\calL^{(\ell)}(\hat{t}_j)| < \hat{t}^{(\ell)}_j$, we compute 
	\begin{equation}
	\hat{t}^{(\ell+1)}_j = \frac{\hat{t}^{(\ell)}_j - |\calL^{(\ell)}(\hat{t}_j)|}{2}.
	\end{equation}
				Then, if $\hat{t}^{(\ell+1)}_j \not \in \ZZ$, we return an error; otherwise, we compute
	\begin{align}
	\by^{(\ell+1)}_j &= \frac{\by^{(\ell)}_j - \sum_{u \in \calL^{(\ell)}(\hat{t}_j)} \bx(u)}{2} \label{eq:recursionmethod}
	\end{align}
	and $\calL^{(\ell+1)}(\hat{t}_j) = \Phi(\by^{(\ell+1)}, \hat{t}^{(\ell+1)}_j)$ and proceed to the next iteration.
	\end{itemize}

It is easy to see that this procedure halts in at most $\tau(\hat{t}_j) \leq \log_2(\hat{t}_j) + 1$ iterations, since $\hat{t}^{(\ell+1)}_j \leq \hat{t}^{(\ell)}_j/2$ and $\hat{t}^{(\tau(\hat{t}_j))}_j \geq 1$.

Then, for each $\hat{t}_j$ that did not return an error, we compute
\begin{equation}
\label{output}
\calL_j(\hat{t}_j) = \calL^{(1)}(\hat{t}_j) \setminus \cup_{\ell = 2}^{\tau(\hat{t}_j)} \calL^{(\ell)}(\hat{t}_j) 
\end{equation}
and
\begin{align}
\label{eq:noiseestimation}
\hat{\bz}_j(\hat{t}_j) = \by_j - \sum_{\ell=1}^{\tau(\hat{t}_j)} 2^{(\ell - 1)} \sum_{u \in \calL^{(\ell)}(\hat{t}_j) }  \bx(u)
\end{align}
which are meant to be used exactly as described in the original OP decoding.

{Note that there is some probability that the ICR method fails, even when we are able to successfully decode the first iteration, since on each subsequent iteration a new decoding attempt has to be performed for the code $\calC$. In order to include these error events, let us redefine the event $E_2$ as follows.} 
{We now denote by $E_2$ the event that the receiver is unable to successfully decode $\calC$ over a bi-AWGN mod 2 channel under noise $\tilde{\bz}_j/2^{\ell}$ (see \eqref{eq:effective-channel-CoF}), for some $\ell \in \{0, 1, 2, \ldots, \tau(t_j) - 1\}$.}
{We compute an approximation for the probability of this event at the end of this section.}

The correctness of our decoding procedure is shown in the following result.

\begin{theorem}
	\label{thm:list}
	If $E_1$ and $E_2$ do not occur, then, the following hold:
	\begin{enumerate}[(i)]
		\item the list $\calL_j(t_j)$ contains all the indices that were transmitted only once in the sub-block $j$;
		\item $\hat{\bz}_j(t_j) = \bz_j$. \label{thm:noise}
	\end{enumerate} 
\end{theorem}

{The proof of the theorem can be found in Appendix~\ref{ssec:proofs}.

To demonstrate the idea of the method and to help understand the notation, we provide a simple example below}.

\begin{example}
	Consider the scenario where the users 
		$\calA_j = \{1, 2, 3, 4, 5, 6, 7, 8\}$ 
	transmitted the indices $u_1$, $u_2$, $u_3$, $u_1$, $u_1$, $u_2$, $u_2$, $u_2$ (where $u_1$, $u_2$, $u_3$ are distinct), respectively, in the sub-block $j$ and $T \geq 8$. In other words, $u_1 = u_4 = u_5$ and $u_2 = u_6 = u_7 = u_8$.
	
	The received signal can then be written as
	\begin{align}
	\mathbf{y}_j &= \sum_{i \in \calA_j} \bx(u_i) + \mathbf{z}_j \nonumber\\&= 3 \bx(u_1) + 4\bx(u_2) + \bx(u_3) + \bz_j.
	\end{align}
	Assuming $\hat{t}_j = 8$, the computation of \eqref{CoFcomputation} for $\mathbf{y}_j^{(1)} = \mathbf{y}_j$ and $\hat{t}_j^{(1)} = \hat{t}_j$ gives
	\begin{align}
	\tilde{\by}_j^{(1)} = \left[ \bc_1 + \bc_3 + \tilde{\z}_j \right] \bmod 2.
	\end{align}
	Assuming we are able to successfully decode $\calC$, then we are able to correctly recover $\calL^{(1)}(\hat{t}_j) = \{u_1, u_3\}$, since $2 \leq T$. This allows us to reconstruct $\bx(u_1)$ and $\bx(u_3)$. However, $2 = |\calL^{(1)}(\hat{t}_j)| < \hat{t}_j^{(1)} = 8$. Thus, we compute
	\begin{align}
	\y^{(2)}_j &= \frac{\left(\y^{(1)}_j - \bx(u_1) - \bx(u_3) \right)}{2}\nonumber\\ &= \bx(u_1) + 2\bx(u_2) + \frac{\z_j}{2}
	\end{align}
	and repeat the decoding algorithm for $\hat{t}_j^{(2)} = \frac{8 - 2}{2} = 3$, recovering $\calL^{(2)}(\hat{t}_j) = \{ u_1\}$. Again, $|\calL^{(2)}(\hat{t}_j)| < \hat{t}_j^{(2)}$, so we compute
	\begin{align}
	\by^{(3)}_j &= \frac{\left(\y^{(2)}_j - \bx(u_1) \right)}{2} = \bx(u_2) + \frac{\bz_j}{4}
	\end{align}
	and repeat the decoding algorithm for $\hat{t}_j^{(3)} = \frac{3-1}{2}=1$, recovering $\calL^{(3)}(\hat{t}_j) = \{u_2\}$. Now, $|\calL^{(3)}(\hat{t}_j)| = \hat{t}_j^{(3)}$, thus the method returns all the computed lists.
	
	Finally, we compute $\calL_j(\hat{t}_j) = \calL^{(1)}(\hat{t}_j) \setminus (\calL^{(2)}(\hat{t}_j) \cup \calL^{(3)}(\hat{t}_j)) = \{u_3\}$
	and
	\begin{align}
	\hat{\bz}_j(\hat{t}_j) &= \by_j - \bx(u_1) - \bx(u_3) - 2 \bx(u_1) - 4\bx(u_2) \nonumber \\&= \bz_j.
	\end{align}
	{We can then check that $\hat{\bz}_j(\hat{t}_j)$ is Gaussian, thus $\hat{t}_j = 8$ is probably correct}. 
\end{example}

Now, note that, if the method succeeds, an error occurs in the scheduling phase only for the users that picked the same index, in contrast to all users in the same sub-block. Therefore, the probability of error presented in \eqref{ep3OP} changes to
\begin{equation}
\epsilon_3  \triangleq 1 - \left(1 - \frac{1}{V n_p} \right)^{K_a - 1} \label{ep3}
\end{equation} 
since this error event is now expressed as \mbox{$E_3 = \{S_{v_i, u_i} > 1\}$}.

{Finally, as we mentioned earlier, the error event $E_2$ was redefined and, therefore, its probability also changes. We can apply the union bound to the events that make up $E_2$ and approximate the probability of error as

\begin{multline}
\epsilon_2 \triangleq \sum_{\ell = 1}^{\log_2(T) + 1} Q\left(\left(C_{\textrm{mod2}}(2^{\ell-1} P_1 V) - \frac{m_p}{n_{c, 1}}\right) \right. \\
\cdot \left. \sqrt{\frac{n_{c, 1}}{\textsf{V}_{\textrm{mod2}}(2^{\ell-1} P_1 V)}}\right)
\end{multline}
where the sum goes up to $\log_2(T) + 1$ as a worst case scenario of $\tau(t_j) \leq \log_2(t_j) + 1$.}

\subsection{Opportunistic Message Transmission}

In the scheme as presented, in the scheduling request phase, the users choose, randomly and uniformly, an index $u$ and a sub-block $v$ to transmit. However, the scheduling request phase may be used opportunistically to transmit part of the message $w_i$, which can be done by choosing the pair $(v, u)$ based on the message. This improvement, which we call Opportunistic Message Transmission (OMT), is described more precisely as follows.

Let $k_1 = \lfloor\log_2 (V \cdot n_p)\rfloor$. Each user $i$ takes the first $k_1$ bits of the message and converts it to its decimal representation (plus 1), denoted by $s_i' \in \{1, 2, \ldots, 2^{k_1}\}$. Then, if $s_i' \in \{1, 2, \ldots, V n_p - 2^{k_1}\}$, the user chooses randomly $s_i \in \{s_i', s_i' + 2^{k_1}\}$. Otherwise, $s_i = s_i'$. Finally, each value of $s_i \in \{1, 2, \ldots, V n_p\}$ is mapped into a pair $(v_i, u_i)$, such that $s_i = v_i + V (u_i - 1)$. Note that the solution is unique, i.e., there is only one pair that satisfies this equation. Conversely, each pair is mapped into a unique $s_i$, which can be mapped back into a unique $s_i'$, i.e., if we are able to recover the pair $(v_i, u_i)$, we are able to recover $s_i'$ and therefore the $k_1$ bits of the message $w_i$.

Previously, each pair $(v, u)$ had a uniform distribution with probability $\frac{1}{V n_p}$, however, this no longer holds. Pairs such that $s_i' \leq V n_p - 2^{k_1}$ are chosen with probability $\frac{1}{2^{k_1+1}}$, while pairs such that $s_i' > V n_p - 2^{k_1}$ are chosen with probability $\frac{1}{2^{k_1}}$. For example, consider $V = 3$ and $n_p = 4$, then $k_1 = 3$. The probability of each position $(v, u)$ is
\begin{equation*}
\textrm{Pr}[(v, u)] = \begin{pmatrix}
1/16 & 1/16  &1/8  &1/16 \\ 
1/16 & 1/8  &1/8  &1/16 \\ 
1/16 & 1/8  &1/16  &1/16\\ 
\end{pmatrix}.
\end{equation*}
This changes the probability that a user chooses a sub-block $v$, which is given by $\textrm{Pr}[v_i = v] = \sum_{u} \textrm{Pr}[v_i = v, u_i = u]$. 

As a bound, we consider the case with highest probability, i.e., we denote by $\textrm{Pr}[v_i = v^*] = \max_{v} \textrm{Pr}[v_i = v]$. In the above example, this corresponds to the second line, which has probability $3/8$, while the first and last have a probability of $5/16$. This probability is given by
\begin{equation}
p(v^*) \triangleq \textrm{Pr}[v_i = v^*] = 
\frac{n_p - \lfloor 2 \frac{V n_p - 2^{k_1}}{V} \rfloor}{2^{k_1}} + \frac{\lfloor 2 \frac{V n_p - 2^{k_1}}{V} \rfloor}{2^{k_1 + 1}}
\end{equation}
This changes \eqref{ep1} to
\begin{align}
\epsilon_1 \triangleq 1 - \sum_{t = 0}^{T - 1} \binom{K_a - 1}{t} \left(p(v^*)\right)^t \left(1 - p(v^*)\right)^{K_a - 1 - t}. \label{ep1_new}
\end{align}

Similarly, since some pairs are more likely than others, \eqref{ep3} changes to
\begin{align}
\epsilon_3   \triangleq& \textrm{Pr}[s_i' \leq V n_p - 2^{k_1}]\cdot \textrm{Pr}[E_3 | s_i' \leq V n_p - 2^{k_1}] \nonumber\\
&+ \textrm{Pr}[s_i' > V n_p - 2^{k_1}] \cdot \textrm{Pr}[E_3 | s_i' > V n_p - 2^{k_1}] \nonumber \\
= &\left(\frac{Vn_p - 2^{k_1}}{2^{k_1}}\right) \cdot \left(1 - \left(1 - \frac{1}{2^{k_1+1}} \right)^{K_a - 1} \right) \nonumber\\
&+ \left(\frac{2^{k_1+1} - Vn_p}{2^{k_1}}\right) \cdot \left(1 - \left(1 - \frac{1}{2^{k_1}} \right)^{K_a - 1} \right).\label{ep3_new}
\end{align}

On the other hand, we now need to transmit only $k_2 = k - k_1$ bits in the data transmission phase. Therefore, \eqref{ep4} changes to 
\begin{equation}
\epsilon_4 \triangleq Q\left(  \left(C_{\textrm{AWGN}}(P_2 K_a) - \frac{k_2}{n_{c, 2}}\right) \frac{\sqrt{n_{c, 2}}}{\sqrt{\textsf{V}_{\textrm{AWGN}}(P_2 K_a)}} \right). \label{ep4_new} 
\end{equation}

\section{Optimization and results}\label{Results}

In this section, we briefly present our optimization framework and the simulation results achieved by our method. 
\subsection{Optimization}

Recall that, given $k$ and $K_a$, our scheme depends on the parameters $k_p$, $n_p$, $n_{c, 1}$, $n_{c, 2}$, $V$, $N_f$, $P_1$, $P_2$ and $P_f$. 
We wish to choose these parameters in order to minimize $E_b/N_0$, subject to constraints on $\epsilon$ and $N$. {Note that, due to the large number of parameters, all of which are linked through the constraints on $N$ and $\epsilon$, individually analyzing the effects of each parameter in the performance is not easy. Furthermore, we have derived approximations for most of the probabilities of error, which strongly motivates us to employ numerical optimization methods in order to find {a good choice of} parameters.}
{For this design, since there are no known bounds or approximations on $\epsilon_e$ and $\epsilon_G$, these probabilities are not considered in the optimization. However, in order to prevent them from growing too large, we add a constraint $n_{c, 1} \geq n_{c, 1}^{\textrm{min}}$, as discussed in Section~\ref{erroranalysis}.} Additionally, in order to simplify the optimization, we use\footnote{For the scenarios considered here, this simplification is found experimentally to have negligible impact on performance.}\textsuperscript{,}\footnote{While at first glance $P_f = P K_a$ might seem large, note that this is exactly the same instantaneous power $P_2 K_a = P K_a$ of the users during their respective sub-blocks in the data transmission phase, thus it is perfectly reasonable to assume that the base station can transmit at this power.} $P_1 = P_2 = P_f/K_a = P$. In this case, since $N$ and $k$ are fixed, minimizing $E_b/N_0 = \frac{PN}{2k}$ is equivalent to minimizing $P$.

This optimization is {still} hard in general since all variables, except $P$, are integers. Thus, we follow a heuristic approach. Given a fixed $k_p \in \ZZ$, $n_p \in \ZZ$, we find the maximum possible minimum distance for $\calC_{\textrm{aux}}$ with these parameters, obtaining $T$. Note that this is only computationally feasible because these parameters are chosen to be small. Then, we relax the integer constraints on the remaining variables, finding their (locally) optimal values over $\RR$ given suitably chosen initial points (we use the default interior-point algorithm on MATLAB's function \texttt{fmincon}). This is possible since all the error expressions are computable with real variables (except $\epsilon_1$ with $T$). The resulting value of $P$ gives a lower bound on the achievable $P$. Then, we fix $n_{c, 1}$, using its optimal real value rounded to the nearest integer and clipped from below at $n_{c, 1}^{\textrm{min}} = 10$, and we run the relaxed optimization for the remaining variables. Then, for each variable in the sequence $n_{c,2}$, $V$, its optimal relaxed value is rounded to the nearest integer and kept fixed, followed by a new run of the relaxed optimization for the remaining variables. After $n_{c,2}$ and $V$ are fixed, we choose $N_f = N - N_1 - N_2$ and compute the required power $P$ to achieve the target error $\epsilon$. This process gives an approximated $P$ for the initially chosen pair $(k_p, n_p)$. Finally, this process is repeated for $k_p = 1, \ldots, k_p^{\textrm{max}}, n_p = k_p+1, \ldots, n_p^{\textrm{max}}$ to find the smallest $P$. While this approach is not guaranteed to be optimal, it has shown experimentally to yield values of $P$ very close to the relaxed lower bound.

As in \cite{Or2017,Vem2017,CCS2018,Pradhan2019}, we use $k = 100$, $N = 30000$ and $\epsilon = 0.05$. The parameters obtained through the optimization process are presented in Table~\ref{table:prob1}. 
{The resulting performance of our scheme, using the same error probability approximations, is shown in Fig.~\ref{fig:ebn0} (curve labeled ``Proposed (Approx.)''). To illustrate the usefulness of ICR and OMT, we also show the approximate performance of our scheme without ICR and OMT, and with ICR but without OMT (labeled as ``Proposed w/o OMT (Approx.)'').}

\subsection{Simulation Results} 

{In order to validate our approximations, we simulate our scheme using the parameters in Table~\ref{table:prob1}.} 
{As usual in the literature, we consider a deterministic, known $K_a$.} For the resource distribution and data transmission phases, we use the error probabilities presented in Section~\ref{erroranalysis}, as these bounds are theoretically achievable and can be approached in practice with reasonable complexity \cite{Liva2016}. For the scheduling request phase, we simulate the performance using the decoding scheme described in the Appendix. {In particular, note, from Table~\ref{table:prob1}, that the code lengths in the scheduling request are remarkably small. This allows us to apply maximum-likelihood decoding for $\calC$ and bounded-distance syndrome decoding for $\calC_{\textrm{aux}}$. Alternatively, ordered statistics decoding \cite{OSD} may be used to achieve near-ML performance.} {We emphasize that this simulation includes all the probabilities of error, except for $\epsilon_4$, $\epsilon_{f, 2}$ and $\epsilon_f$, which are proven achievable bounds. Especially, we note that $\epsilon_e$, $\epsilon_{2}$ and $\epsilon_G$
{(for which we have previously used approximations not proven to be achievable)}
are simulated.} Finally, if the error probability is smaller than the target $\epsilon$, we decrease $P_2$ until the error probability is achieved with equality. On the other hand, if the simulated error probability is larger than desired, we increase $P_2$.

\begin{figure}
	\centering
	\includegraphics[scale=.63]{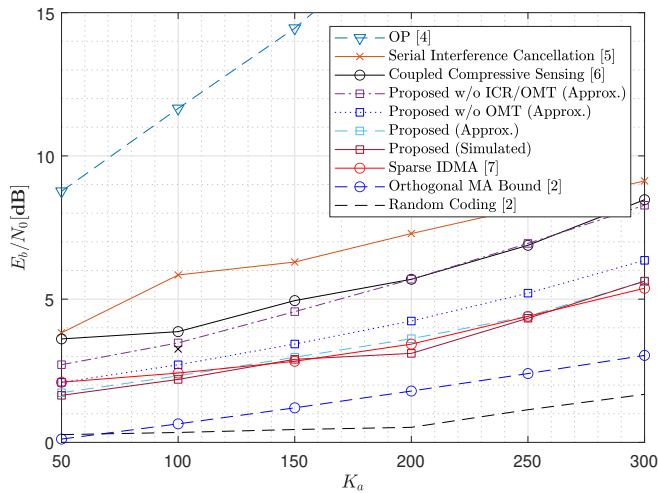}
	\caption{Comparison between the $E_b/N_0$ required for $k = 100$ bits, $N = 30000$ channel uses, $\epsilon = 0.05$.}
	\label{fig:ebn0}
\end{figure}

The results are presented in Fig~\ref{fig:ebn0} {(curve labeled ``Proposed (Simulated)'')} and compared to \cite{Or2017, Vem2017, CCS2018,Pradhan2019}. {The {data} points for these works are extracted directly from their papers.}
We also plot the random coding bound and the orthogonal MA bound \cite{Polyanskiy2017}, where perfect coordination is assumed a priori. Note that, since we use orthogonal methods in the data transmission, the latter provides a lower bound to our method. As can be seen, our method performs closely to the state-of-art in \cite{Pradhan2019}. Additionally, note that our approximations {for the scheduling phase} seem to be tight in this regime of operation, with the performance of the simulation being close to the approximation. 

{Finally, as a case study, for $K_a = 100$, we include a simulation using variable-regular binary LDPC codes for both the feedback and the data transmission phases, assuming binary antipodal modulation. Both codes are constructed using the progressive edge-growth algorithm \cite{Hu.etal.2005.Regular-Irregular-Progressive} with a variable-node degree of 3 and are decoded using belief propagation. Since such codes do not achieve the finite blocklength achievability previously used, we increase the powers $P_2$ and $P_f$ in order to achieve the desired probability of error.} 
{The result is shown in the point marked as ``X'' in Fig.~1. As can be seen, there is a gap of approximately $1$~dB resulting from the use of such codes instead of the finite blocklength achievability.
}
{A better result could potentially be achieved by using state-of-the-art short-blocklength codes that closely approach the fundamental limits \cite{Coskun.etal.2019.Efficient-Error-correcting-Codes}, designed for the specific parameters in Table~\ref{table:prob1}.}

\begin{table*}[]
\centering
\caption{Optimized parameters for the proposed scheme with ICR and OMP}
\begin{tabular}{|c|c|c|c|c|c|c|}
\hline
$K_a$        & $50$  & $100$ & $150$ & $200$ & $250$ & $300$ \\ \hline
$k_p$           & $5$   & $5$   & $5$   & $5$   & $5$   & $13$   \\ \hline
$n_p$         & $15$   & $15$   & $15$   & $15$   & $15$   & $31$   \\ \hline
$d$ &$7$ &$7$ &$7$ &$7$ &$7$ & $9$ \\ \hline
$n_{c, 1}$ & $18$  & $17$  & $14$  & $13$  & $12$  & $21$  \\ \hline
$n_{c, 2}$ & $402$ & $210$ & $142$ & $109$  & $88$  & $70$  \\ \hline 
$V$          & $448$ & $420$ & $488$ & $518$ & $548$ & $362$ \\ \hline
$N_f$ & $1836$ & $1860$ & $1868$ & $1466$ & $1424$ & $1398$ \\ \hline
$P$ & $0.009936$ & $0.01137$ & $0.01320$ & $ 0.01532$ & $0.01837$ & $0.02434$\\ \hline
\end{tabular}
\label{table:prob1}
\end{table*}

\section{Discussion}

The proposed model is idealized in that it assumes perfect synchronization, absence of fading, absence of processing delays, and knowledge of $K_a$ by the base station and by the users (which, by the very nature of the random access problem, is generally unknown). These assumptions stem from Polyanskiy's model, of which our proposed model is an extension. To our understanding, Polyanskiy's model is an intentionally simplified model that captures essential aspects of massive random access, in order to allow a first step towards a fundamental understanding of the problem. Naturally, further work is needed addressing more realistic scenarios.\footnote{Actually, absence of fading may not be entirely unrealistic if users apply channel inversion after receiving a pilot signal from the base station, as in \cite{RaptorM2M}.}

The purpose of this paper is to provide a comparison that is as fair as possible between our proposed grant-based scheme and existing grantless schemes under Polyanskiy's model. In this sense, the practical issues mentioned above do not affect the main conclusions of this work. Indeed, with respect to fading, our proposed scheme is arguably more robust, as this issue can be better mitigated or more easily handled in the coordinated data transmission phase. 

On the other hand, while the proposed scheme certainly will have a larger processing delay due to the two additional phases, such a delay is likely to be much smaller than the duration of a session (which aggregates the transmissions of $K_a$ users) and therefore have a minor impact on performance. Note that this delay refers only to signal/data processing; the transmission delay is already fixed by the total number of channel uses $N$.

{Additionally}, we should mention that the exact knowledge of $K_a$ is not critical to our scheme and may be replaced by a base station's estimate of $K_a$ based on past sessions. This is possible since the performance of the scheme gracefully degrades with the estimation error (an overestimate increases the energy consumption, while an underestimate increases the error probability) and both situations can be detected by the base station after the scheduling request.

{While our best results assume that codes achieving the normal approximation for the AWGN channel \eqref{AWGN} are used in the feedback and data transmission phases, these codes may also be replaced by suboptimal ones, albeit with a corresponding energy penalty. The design of efficient short-blocklength codes is an active area of research \cite{Coskun.etal.2019.Efficient-Error-correcting-Codes},} and, as practical codes that approach the bounds are developed, our results will also be easier to achieve with low complexity.

\section{Conclusion}\label{Conclusion}

We presented a new grant-based scheme for the massive random access Gaussian channel which allows coordinating the users for a better overall performance at the low cost of a short broadcast feedback from the base station. Such a cost, both in terms of energy consumption and transmission delay, is accounted for in the model to allow for a fair comparison with grantless schemes. The scheduling strategy we use is based on channel codes and has the advantage that its error probability can be approximated by a closed-form expression, allowing a simple design and optimization of the scheme. Simulation results show that our method performs closely to the state-of-art in \cite{Pradhan2019}, outperforming previous practical grantless schemes in \cite{Or2017, Vem2017} and \cite{CCS2018}. Overall, our results suggest that novel grant-based schemes specifically designed for a massive number of users may be a competitive solution for next-generation random access applications.

\appendix

\subsection{Original OP Decoding \cite{Or2017}}
\label{ssec:original-op}

Recall that, for each sub-block, the base station receives $\by_j$, described in \eqref{receivedUP1}.

Let $\hat{t}_j$ be a receiver's estimate of $t_j = |\mathcal{A}_j|$. The receiver computes
\begin{align}
\tilde{\by}_j \triangleq \left[ \frac{1}{2\sqrt{P_1V} }\y_j + \frac{\hat{t}_j}{2}  \right]\bmod 2 \label{CoFcomputation}
\end{align}
where the modulo 2 reduction is into the interval [0, 2) and is taken componentwise. 

If $\hat{t}_j = t_j$, then
\begin{align}
\tilde{\by}_j = \left[ \sum_{i \in \mathcal{A}_j } \bc_{i} + \tilde{\z}_j  \right] \bmod 2 \label{CoF}
\end{align}
where $\tilde{\z}_j = \frac{\z_j}{2 \sqrt{P_1V}}$. Note that, since the code $\calC$ is linear, the sum of codewords $\tilde{\bc}_j \triangleq \sum_{i \in \mathcal{A}_j }\bc_{i}\bmod 2$ also belongs to the code $\calC$. Thus, the effective channel is
\begin{align}
\label{eq:effective-channel-CoF}
\tilde{\by}_j = (\tilde{\bc}_j + \tilde{\bz}_j)\bmod 2.
\end{align}
The codeword $\tilde{\bc}_j$ is decoded through maximum-likelihood decoding and, since $\tilde{\bc}_j = \left(\sum_{i \in \mathcal{A}_j } \bh_{u_i} \right)\bG\bmod 2$, if the decoding is successful, we recover
\begin{align}
\tilde{\bh}_j \triangleq \sum_{i\in \mathcal{A}_j} \bh_{u_i} \bmod 2.
\end{align}
Note that $\tilde{\bh}_j$ can be viewed as a syndrome of $\calC_{\textrm{aux}}$ over the binary symmetric channel, corresponding to ``errors'' in positions $u_i$, $i \in \calA_j$.
Then, a bounded-distance syndrome decoder is applied to $\tilde{\bh}_j$, producing a list $\mathcal{L}(\hat{t}_j)$ of estimated indices $u_i$. Due to the error-correction capability of $\calC_{\textrm{aux}}$, this list will be exactly $(u_i, i \in \calA_j)$ if all the $u_i$'s are distinct and $|\calA_j| \leq T$.

On the other hand, if $\hat{t}_j \neq t_j$, the computation of $\tilde{\by}_j$ results in
\begin{align}
\tilde{\by}_j = \left[ \sum_{i \in \mathcal{A}_j } c_{i} + (\hat{t}_j - t_j) + \tilde{\z}_j  \right] \bmod 2 \label{CoF_error}.
\end{align}
which will likely cause an error in the decoding of the linear code if $(\hat{t}_j - t_j)\bmod 2 \neq 0$. If an undetected error occurs on this step, a wrong $\hat{\tilde{\bh}}_j \neq \tilde{\bh}_j$ is returned, which will be then decoded by $\calC_{\textrm{aux}}$, generating a list $\mathcal{{L}}(\hat{t}_j)$ of estimated indices. 

For any $\hat{t}_j$, if a detected error occurs in any decoding step, the output is set to $\calL(\hat{t}_j) = \emptyset$ and an error is flagged.

Let this decoding procedure, which depends on $\by_j$ and the estimated $\hat{t}_j$, be denoted by a function $\calL(\hat{t}_j) =  \Phi(\by_j, \hat{t}_j)$. We compute $\calL(\hat{t}_j) =  \Phi(\by_j, \hat{t}_j)$ for $0 \leq \hat{t}_j \leq T$, generating $T+1$ lists $\calL(\hat{t}_j)$. For each list, the base station can regenerate an estimate of the transmitted signals ${\bx}(u)$ based on $\calL(\hat{t}_j)$ and subtract them from $\by_j$, generating $\hat{\bz}_j(\hat{t}_j) = \by_j - \sum_{u \in \calL(\hat{t}_j)} {\bx}(u)$. For $\hat{t}_j = t_j$, if no error has occurred in any step of the decoding, this yields $\hat{\bz}_j(\hat{t}_j) = \bz_j$. For $\hat{t}_j \neq t_j$, if no {undetected} error has occurred, this results in $\hat{\bz}_j(\hat{t}_j) = \bz_j + \sum \bx$, where $\sum \bx$ is some unknown sum of transmitted signals. Only in the former case the components of $\hat{\bz}_j(\hat{t}_j)$ are i.i.d.\ samples of a Gaussian random variable. Therefore, as argued in \cite{Or2017}, the base station can choose the  $\hat{t}^*_j$ which yields the best agreement with $\by_j$ and set $\calL_j = \calL(\hat{t}^*_j)$. Specifically, we choose the $\hat{t}_j$ that minimizes the quadratic norm of $\hat{\bz}_j(\hat{t}_j)$.

\subsection{Proof of Theorem~\ref{thm:list}}\label{ssec:proofs}

\newcommand{\bzl}{{\bz}^{(\ell)}_j}
\newcommand{\tbzl}{\tilde{\bz}^{(\ell)}_j}

Let us introduce some notation. We denote by $\calA^{(1)}_j(u) \subseteq \calA_j$ the set of users that transmitted the index $u$ in the sub-block $j$, i.e., if $i \in \calA^{(1)}_j(u)$, then $\bx_{i,j} = \bx(u)$. Additionally, for $\ell \geq 1$, let $\calB^{(\ell)}_j(u) \subseteq \calA^{(\ell)}_j(u)$ be some subset such that \[|\calB_j^{(\ell)}(u)| = 2 \left\lfloor \frac{|\mathcal{A}_j^{(\ell)}(u)|}{2} \right\rfloor\] 
let $\calA^{(\ell+1)}_j(u) \subseteq \calB^{(\ell)}_j(u)$ be some subset such that 
\[|\calA^{(\ell+1)}_j(u)| = \frac{|\calB^{(\ell)}_j(u)|}{2}\] 
and let
\begin{align}
\calB^{(\ell)}_j &\triangleq \cup_{u \in \{1, 2, \ldots, n_p\} } \calB^{(\ell)}_j(u)\\
\calA^{(\ell)}_j &\triangleq \cup_{u \in \{1, 2, \ldots, n_p \} } \calA^{(\ell)}_j(u).
\end{align}
Finally, we define $\calU(\calS) \triangleq \{u_i, i \in \calS  \}$, where $\calS \subseteq \{1, 2, \ldots, K_a\}$ is some set of users.

For example, if the users $\calA^{(1)}_j = \{2, 3, 9, 12\}$ transmitted the corresponding indices $\{7, 7, 7, 4\}$ in the $j$-th sub-block, then $\calA^{(1)}_j(7) = \{2, 3, 9\}$ and $\calA^{(1)}_j(4) = \{12\}$. Some of the possible subsets are $\calB^{(1)}_j(7) = \calB^{(1)}_j = \{2, 3\}$ and $\calA^{(2)}_j(7) = \calA^{(2)}_j = \{2\}$. We also have $\calU\left(\calA_j^{(1)} \right) = \{7, 4\}$ and $\calU(\calA^{(2)}_j ) = \calU(\calB^{(1)}_j )= \{7\}$.

Note that, by construction, $|\calA^{(\ell)}_j(u) \setminus \calB^{(\ell)}_j(u)|$ is either 0 or 1, therefore, $\left|\calU\left(\calA^{(\ell)}_j \setminus \calB^{(\ell)}_j  \right)\right| = |\calA^{(\ell)}_j \setminus \calB^{(\ell)}_j |$ for all $\ell \geq 1$. It is also easy to see that $\calU(\calA^{(\ell+1)}_j ) = \calU(\calB^{(\ell)}_j )$. This will be useful later.

To simplify notation, let $\tau = \tau(t_j)$ for the remainder of this subsection.

\begin{lemma}
	If $E_1$ and $E_2$ do not occur and $\hat{t}_j = t_j$, then for all $\ell \geq 1$, the following hold:
	\begin{enumerate}[(i)]
		\item $\hat{t}^{(\ell)} = | \calA_j^{(\ell)}|$;
		\item $\by_j^{(\ell)} = \sum_{i \in \calA^{(\ell)}_j} \bx_{i,j} + \bz_j/2^{\ell-1}$;
		\item The decoding of the $\ell$th iteration is successful;
		\item $\calL^{(\ell)}(\hat{t}_j) = \calU\left(\calA^{(\ell)}_j\setminus \calB^{(\ell)}_j\right)$.
	\end{enumerate}
	Moreover, $\calB_j^{(\tau)} = \emptyset$, which implies $\calL^{(\tau)}(\hat{t}_j) = \calU\left(\calA^{(\tau)}_j\right)$.
\end{lemma}
\begin{IEEEproof}
	First, we prove the main statement for $\ell = 1$. Part (i) follows from the assumption $\hat{t}_j = t_j$, since $\hat{t}^{(1)}_j = \hat{t}_j$ and $t_j = |\calA_j| = |\calA^{(1)}_j|$ by definition. Part (ii) also follows directly from definition. Part (iii) follows from the assumptions.
	
	In order to prove (iv), we write the received signal as
	\begin{align}
	\y^{(1)}_j = \by_j = \sum_{i \in \calB^{(1)}_j} \x_{i,j} + \sum_{i \in \mathcal{A}^{(1)}_j \setminus \calB^{(1)}_j } \x_{i,j} + \z_j.
	\end{align}
	Since $\hat{t}^{(1)}_j = t_j$, the computation of \eqref{CoFcomputation} yields
	\begin{align}
	\tilde{\by}_j^{(1)} &= \left[\sum_{i \in \calB^{(1)}_j}\bc_i + \sum_{i \in \mathcal{A}^{(1)}_j \setminus \calB^{(1)}_j } \bc_i + \tilde{\bz}_j    \right] \bmod 2 \\
	&= \left[ \sum_{i \in \mathcal{A}^{(1)}_j \setminus \calB^{(1)}_j } \bc_i + \tilde{\bz}_j    \right] \bmod 2 \label{CoF_1}
	\end{align}
	where \eqref{CoF_1} follows from $|\calB^{(1)}_j(u)|$ being even for all $u$ by construction, which implies that the sum $\sum_{i \in \calB^{(1)}_j}\bc_i$ vanishes.
	
	Since $\calC$ is assumed to successfully decode this sum of codewords {(i.e., $E_2$ is assumed to not occur)} and $\calC_{\textrm{aux}}$ is assumed to successfully recover all the indices transmitted by the users in $\calA^{(1)}_j\setminus \calB^{(1)}_j$ {(i.e., $E_1$ is assumed to not occur)}, then $\calL^{(1)}(\hat{t}_j) = \calU\left(\calA^{(1)}_j\setminus \calB^{(1)}_j\right)$.
	
	We now show that, if the lemma holds for $\ell$, it also holds for $\ell+1$.
	
	Since the lemma holds for $\ell$, i.e., $\hat{t}^{(\ell)}_j = |\calA^{(\ell)}_j|$ and $|\calL^{(\ell)}(\hat{t}_j)| = |\calA^{(\ell)}_j\setminus\calB^{(\ell)}_j|$, then we have  $\hat{t}^{(\ell+1)}_j = \frac{|\calA^{(\ell)}_j| - |\calA^{(\ell)}_j\setminus\calB^{(\ell)}_j|}{2} = \frac{|\calB^{(\ell)}_j|}{2} = |\calA^{(\ell+1)}_j|$, which completes the proof of (i).
	
	The computation of the $(\ell+1)$th iteration is given by
	\begin{align}
	\by^{(\ell+1)}_j &= \frac{\by^{(\ell)}_j - \sum_{u \in \calL^{(\ell)}_j} \bx(u)}{2} \label{eq:proof1_def1}\\
	&= \frac{\by^{(\ell)}_j - \sum_{i \in \mathcal{A}^{(\ell)}_j \setminus \calB^{(\ell)}_j} \bx_{i,j}}{2} \label{eq:proof1_hyp1}\\
	&= \frac{\sum_{i \in \calB^{(\ell)}_j} \bx_{i,j} + \bz_j/2^{\ell-1} }{2}\label{eq:proof1_hyp2} \\
	&= \sum_{u \in \{1, 2, \ldots, n_p\}} \sum_{i \in \calB^{(\ell)}_j(u)} \frac{\bx_{i,j}}{2} + \frac{\bz_j}{2^\ell}\\
	&= \sum_{u \in \{1, 2, \ldots, n_p\}} |\calB^{(\ell)}_j(u)| \frac{\bx(u)}{2} + \frac{\bz_j}{2^\ell} \label{y2}\\
	&= \sum_{u \in \{1, 2, \ldots, n_p\}} |\calA^{(\ell+1)}_j(u)| \bx(u) + \frac{\bz_j}{2^\ell}\\
	&=  \sum_{u \in \{1, 2, \ldots, n_p\}} \sum_{i \in \calA^{(\ell+1)}_j(u)} \bx_{i,j} + \frac{\bz_j}{2^\ell}\\
	&= \sum_{i \in \calA^{(\ell+1)}_j} \bx_{i,j} + \frac{\bz_j}{2^\ell}
	\end{align}
	where \eqref{eq:proof1_def1} follows from definition \eqref{eq:recursionmethod}; \eqref{eq:proof1_hyp1} and \eqref{eq:proof1_hyp2} follow from hypothesis; and the remaining follows from the construction of the sets. This completes the proof of (ii).
	
	In order to prove (iii), we rewrite the signal as
	\begin{align}
	\by^{(\ell+1)}_j = \sum_{i \in \calB^{(\ell+1)}_j} \bx_{i,j} + \sum_{i \in \calA^{(\ell+1)}_j \setminus \calB^{(\ell+1)}_j} \bx_{i,j} + {\bz}^{(\ell+1)}_j
	\end{align}
	since $\calB^{(\ell+1)}_j$ is a subset of $\calA^{(\ell+1)}_j$.
	
	Since $\hat{t}^{(\ell+1)}_j  = |\calA^{(\ell+1)}_j|$, the dither is successfully corrected in the $(\ell+1)$th iteration, thus the computation of \eqref{CoFcomputation} for this iteration yields
	\begin{align}
	\tilde{\by}_j^{(\ell+1)} = \left[ \sum_{i \in \mathcal{A}^{(\ell+1)}_j \setminus \calB^{(\ell+1)}_j } \bc_i + \frac{\tilde{\bz}_j}{2^{\ell}}   \right] \bmod 2.
	\end{align}
	{Now, recall that, in Section~\ref{IndexCollision}, we redefined $E_2$ as the event that the decoding of $\calC$ under noise $\tilde{\bz}_j/2^{\ell}$ fails, for some $\ell \in \{1, 2, \ldots, \tau(t_j)-1\}$. Since we assume $E_2$ has not occurred, we are able to decode $\tilde{\by}_j^{(\ell+1)}$ up to $\ell+1 = \tau(t_j)$, that is, up to the last iteration of the method.} 	Additionally, since $ |\mathcal{A}_j| =  t_j \leq T$ and $\calB^{(\ell)}_j \subseteq \mathcal{A}_j$, then $|\calB^{(\ell)}_j|/2 \leq T$, thus the decoding of $\calC_{\textrm{aux}}$ is certainly successful, completing the proof of (iii). This immediately implies that we have $\calL^{(\ell+1)}(\hat{t}_j) = \calU\left(\calA^{(\ell+1)}_j\setminus \calB^{(\ell+1)}_j\right)$, 	completing the proof of (iv).
	
	Note that, under the conditions of the lemma, the method does not return an error, therefore it ends with $|\mathcal{A}^{(\tau)}_j \setminus \calB^{(\tau)}_j| = |\calL^{(\tau)}| = \hat{t}_j^{(\tau)} = |\mathcal{A}^{(\tau)}_j|$, which implies $\calB^{(\tau)}_j = \emptyset$. It immediately follows that $\calL^{(\tau)} = \calU\left(\calA^{(\tau)}_j\setminus \calB^{(\tau)}_j\right) = \calU\left(\calA^{(\tau)}_j\right)$.
\end{IEEEproof}

\begin{corollary}
	If $E_1$ and $E_2$ do not occur and $\hat{t}_j = t_j$, then $\calU\left(\calA^{(\ell)}_j\right)= \cup_{\ell' = \ell}^{\tau} \calL^{(\ell')}(\hat{t}_j)$, for all \mbox{$\ell \geq1$}.
\end{corollary}

\begin{IEEEproof}	
	First, recall that, by construction, $\calU\left(\calA^{(\ell)}_j\right) = \calU\left( \calB^{(\ell - 1))}_j\right)$. Thus, we have
	\begin{align}
	\calU\left(\calA^{(\ell - 1)}_j\right) &= \calU\left(\calB^{(\ell - 1)}_j\right) \cup \calU\left(\calA^{(\ell - 1)}_j\setminus \calB^{(\ell - 1)}_j\right)\\
	&= \calU\left(\calA_j^{(\ell)}\right) \cup \calU\left(\calA^{(\ell - 1)}_j\setminus \calB^{(\ell - 1)}_j\right) \label{eq:corol1_def}
	\end{align}
	for all $\ell \geq 2$. Applying Lemma 1 to \eqref{eq:corol1_def} with $\ell = \tau$ yields
	\begin{align}
	\calU\left(\calA^{(\tau - 1)}_j\right) &= \calU\left(\calA_j^{(\tau)}\right) \cup \calU\left(\calA^{(\tau - 1)}_j\setminus \calB^{(\tau - 1)}_j\right) \\
	&= \calL^{(\tau)}(\hat{t}_j) \cup \calL^{(\tau - 1)}(\hat{t}_j). \label{eq:firstA}
	\end{align}
	
	We can then solve \eqref{eq:corol1_def} recursively, for example, for $\ell = \tau - 1$, we have
	\begin{align}
	\calU\left(\calA^{(\tau - 2)}_j\right)&= \calU\left(\calA_j^{(\tau - 1)}\right) \cup \calU\left(\calA^{(\tau - 2)}_j\setminus \calB^{(\tau - 2)}_j\right) \\
	&= \calL^{(\tau)}(\hat{t}_j) \cup \calL^{(\tau - 1)}(\hat{t}_j) \cup \calL^{(\tau - 2)}(\hat{t}_j).
	\end{align}
	
	Generally, solving the recursion yields
	\begin{align}
	\calU\left(\calA^{(\ell)}_j\right)= \cup_{\ell' = \ell}^{\tau} \calL^{(\ell')}(\hat{t}_j).
	\end{align}
\end{IEEEproof}

\begin{IEEEproof}[Proof of Theorem~\ref{thm:list}]Since $E_1$ and $E_2$ are assumed to not occur and we are considering the case $\hat{t}_j = t_j$, then, from Corollary 1,  
\begin{equation}
\calU\left(\calA^{(\ell)}_j\right)= \cup_{\ell' = \ell}^{\tau} \calL^{(\ell')}(\hat{t}_j).
\end{equation}

It is easy to see that $\calU\left( \calA^{(2)}_j\right)$ contains all the indices that collided, i.e., indices that were transmitted more than once. 
Therefore, the list of all indices that were transmitted only once is given by
\begin{align}
\calU\left( \calA^{(1)}_j\right) \setminus \calU\left( \calA^{(2)}_j\right) &= \left(\cup_{\ell = 1}^{\tau}\calL^{(\ell)}(\hat{t}_j)\right) \setminus \cup_{\ell = 2}^{\tau}\calL^{(\ell)}(\hat{t}_j)\\
&= \calL^{(1)}(\hat{t}_j) \setminus \cup_{\ell = 2}^{\tau}\calL^{(\ell)}(\hat{t}_j).
\end{align} 
Therefore, the output $\calL_j(t_j)$ given in \eqref{output} contains all the indices that have not collided. This completes the proof of (i).

Now, let
\begin{align}
\by^{(\tau+1)}_j &\triangleq \frac{\by^{(\tau)}_j - \sum_{u \in \mathcal{L}^{(\tau)}_j } \bx(u)}{2}\\
&= \frac{\by^{(\tau)}_j - \sum_{i \in \mathcal{A}^{(\tau)}_j} \bx_{i,j}}{2}\\
&= \frac{\bz_j}{2^{\tau}} \label{eq:purenoise}
\end{align}
where the equalities follow immediately from Lemma 1.

Additionally, solving the recursion in $\by^{(\tau)}_j$ using \eqref{eq:recursionmethod}, we also have
\begin{align}
\by^{(\tau+1)}_j &= \frac{\by_j}{2^{\tau}} - \sum_{\ell = 1}^{\tau} \frac{1}{2^{\tau + 1 - \ell}} \sum_{u \in \mathcal{L}^{(\tau)}_j} \bx(u)\\
&= \frac{\hat{\bz}_j(\hat{t}_j)}{2^{\tau}} \label{eq:purenoise2}
\end{align}
where \eqref{eq:purenoise2} follows from the definition in \eqref{eq:noiseestimation}.

Comparing both equations, we have $\hat{\bz}_j(\hat{t}_j) = \bz_j$. This completes the proof of (ii).
\end{IEEEproof}

\begin{IEEEbiography}[{\includegraphics[width=1in,height=1.25in,clip,keepaspectratio]{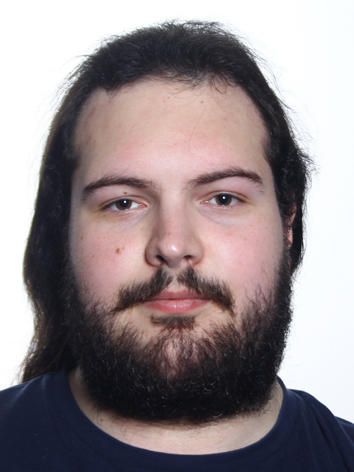}}]
	{Gustavo Kasper Facenda} (S'20) received the B.Sc.\ and the M.Sc. degrees in electrical engineering from the Federal University of Santa Catarina (UFSC), Florian\'opolis, Brazil, in 2017 and 2019, respectively. He is currently a PhD student at University of Toronto, Toronto, Canada. His research interests include information theory and its applications, such as multiple access channels, relay networks and streaming codes.
\end{IEEEbiography}

\begin{IEEEbiography}[{\includegraphics[width=1in,height=1.25in,clip,keepaspectratio]{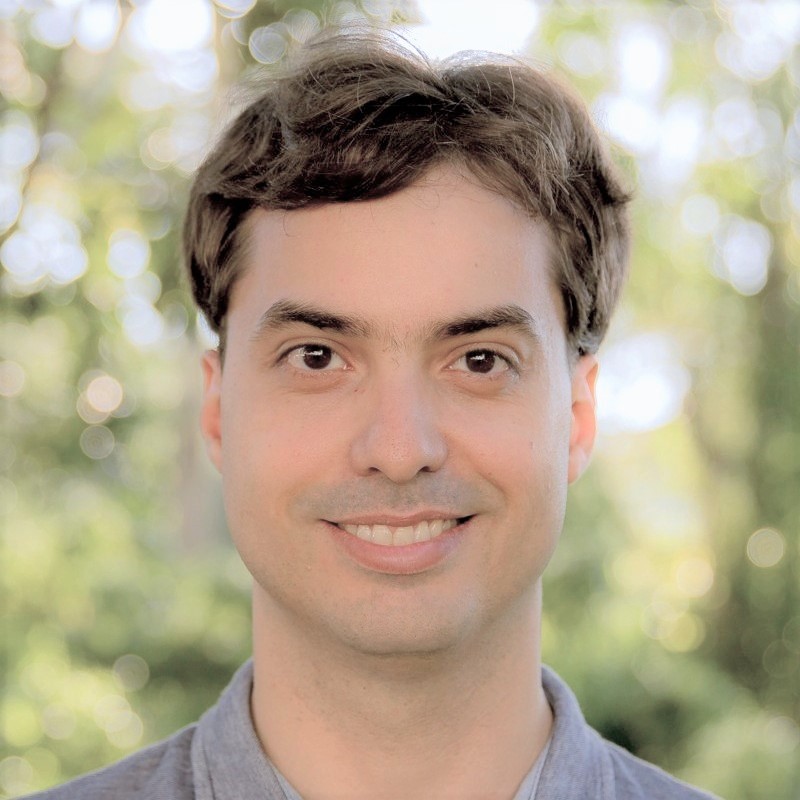}}]
	{Danilo Silva} (S'06--M'09) received the B.Sc.\ degree from the Federal University of Pernambuco (UFPE), Recife, Brazil, in 2002, the M.Sc.\ degree from the Pontifical Catholic University of Rio de Janeiro (PUC-Rio), Rio de Janeiro, Brazil, in 2005, and the Ph.D. degree from the University of Toronto, Toronto, Canada, in 2009, all in electrical engineering.
	
	From 2009 to 2010, he was a Postdoctoral Fellow at the University of Toronto, at the \'Ecole Polytechnique F\'ed\'erale de Lausanne (EPFL), and at the State University of Campinas (UNICAMP).
	In 2010, he joined the Department of Electrical and Electronic Engineering, Federal University of Santa Catarina (UFSC), Brazil, where he is currently an Associate Professor. His research interests include machine learning, information theory, and their applications to signal processing and wireless communications.
	
	Dr. Silva is a member of the Brazilian Telecommunications Society (SBrT). He was a recipient of a CAPES Ph.D. Scholarship in 2005, the Shahid U. H. Qureshi Memorial Scholarship in 2009, and a FAPESP Postdoctoral Scholarship in 2010.
\end{IEEEbiography}

\end{document}